\documentclass[a4paper,12pt]{article}
\usepackage{mathrsfs} \usepackage{amsmath} \usepackage{amssymb} \usepackage{slashed}
\usepackage{graphicx} \usepackage{caption}
\begin{document}
\newcommand{\ba}{\begin{eqnarray}} \newcommand{\ea}{\end{eqnarray}}
\newcommand{\be}{\begin{equation}} \newcommand{\ee}{\end{equation}}
\renewcommand{\figurename}{Figure}
\renewcommand{\thefootnote}{\fnsymbol{footnote}}

\vspace*{1cm}
\begin{center}
 {\Large\textbf{A Model of Neutrino Mass, Baryon Asymmetry, and Asymmetric Dark Matter with $SU(2)_D\otimes U(1)_{D'}$ Dark Sector}}

 \vspace{1cm}
 \textbf{Wei-Min Yang}

 \vspace{0.3cm}
 \emph{Department of Modern Physics, University of Science and Technology of China}

 \emph{Hefei 230026, P. R. China}

 \emph{E-mail: wmyang@ustc.edu.cn}
\end{center}

\vspace{1cm}
\noindent\textbf{Abstract}:
  I suggest a new extension of the standard model of particle physics, which introduces a dark sector with the $SU(2)_{D}\otimes U(1)_{D'}$ symmetry besides the SM sector. The new particles of the model all inhabit in the dark sector. The dark gauge symmetry breaking will bring about fruitful physics beyond the SM. The tiny neutrino mass is generated through the Dirac-type seesaw mechanism. The inflaton decay can not only provide the universe inflation and reheating, but also lead to the baryon asymmetry and the asymmetric cold dark matter. In short, the model provides an unification of the neutrino mass, the baryon asymmetry, the asymmetric CDM and the inflation, and it can account for their common origin. Finally, it is very possible to test the model predictions and probe the dark sector physics in near future experiments.

\vspace{1cm}

\newpage
\noindent\textbf{I. Introduction}

\vspace{0.3cm}
  The standard model (SM) of the fundamental particles has successfully accounted for all kinds of the particle phenomena at or below the electroweak scale, refer to the relevant reviews in Particle Data Group \cite{1}, but it can not address the three important issues: the tiny neutrino mass \cite{2}, the matter-antimatter asymmetry \cite{3}, and the cold dark matter (CDM) \cite{4}. These issues however reveal the new physics beyond the SM, so they are intensively investigated. Particle scientists have established plenty of experimental data for the neutrino physics and the baryon asymmetry \cite{1}, but the CDM has not yet been detected by any one terrestrial experiment except for the evidence from cosmic observations \cite{5}. Any one new theory beyond the SM has to confront these three issues and address them, therefore, searching a correct solution for them may be the largest challenge for theoretical particle physics.

  Many theories have been suggested in the last half century since the SM was established, but a majority of them only focus on one of the issues rather than an integrated solution. The tiny neutrino mass can be generated by the seesaw mechanism \cite{6} or result from some loop-diagram radiative generation \cite{7}. The baryon asymmetry can be achieved by the thermal leptogenesis \cite{8} or the electroweak baryogenesis \cite{9}. The CDM candidates are possibly the sterile neutrino \cite{10}, the lightest supersymmetric particle \cite{11}, the axion \cite{12}, and so on. In recent years, some inspiring ideas attempt to find some connections among the neutrino mass, the baryon asymmetry and the CDM, for instance, the neutrino mass and the leptogenesis can be together implemented by the super-heavy Majorana fermion \cite{13} or the scalar triplet \cite{14}, the asymmetric CDM can be related to the baryon asymmetry \cite{15}, and some sophisticated models can unify them into a framework \cite{16a}, and the author's recent works on this field \cite{16b}. Although many progresses on the new theory beyond the SM have been made all the time, a realistic and convincing theory is not established as yet.

  Based on the universe harmony and the nature unification, it is reasonable and believable that a realistic theory beyond the SM should be able to provide an unification of the tiny neutrino mass, the baryon asymmetry and the CDM, even including the universe inflation and reheating, in other words, all of these things are related to each other and they have a common origin in the new framework. On the other hand, this theory should keep such principles as simplicity, fewer number of parameters, and possibility of being tested in future experiments. If one theory is excessive complexity and/or unable to be tested, it is unbelievable and insignificant. After careful considerations, I suggest a new extension of the SM. The new model introduces a dark sector besides the SM sector, it consists of several species of dark particles and obeys the dark local gauge symmetry of $SU(2)_{D}\times U(1)_{D'}$. The dark gauge symmetry breaking will bring about fruitful physics beyond the SM, in particular, the model can completely account for the common origin of the above-mentioned things. In addition, it is very feasible to test the model and probe the dark sector physics by means of the TeV-scale colliders, the underground detectors, and the cosmic neutrino.

  The remainder of this paper is organized as follows. In Section II, I outline the model. In Section III, I introduce how the inflaton decay leads to the baryon asymmetry and the asymmetric CDM. The dark sector physics is discussed in Section IV. I give some numerical results and discuss the model test in Section V. Section VI is devoted to conclusions.

\vspace{0.6cm}
\noindent\textbf{II. Model with $SU(2)_{D}\otimes U(1)_{D'}$ Dark Sector}

\vspace{0.3cm}
  The minimal extension of the SM is realized by adding three right-handed fermion singlets to the SM, which is commonly called as the $\nu SM$, refer to \cite{16c}. My model introduces a dark sector besides the $\nu SM$ sector, it consists of several species of dark particles which obey the dark local gauge symmetry of $SU(2)_{D}\otimes U(1)_{D'}$. The model particle contents and its symmetries are summarized in Tab. 1, here I omit the color subgroup $SU(3)_{C}$ and the quarks in the SM sector since they are not involved in the following discussions of this paper. Throughout we will signify some physical quantities in the dark sector by the notations with the ``D" superscript or subscript, for example, the dark gauge fields $W_{\mu}^{D},B_{\mu}^{D}$, the dark chiral fermion doublets $l_{L}^{D},l_{R}^{D}$, the dark electric charge $Q^{D}$, and so on.

\begin{table}
\centering
\begin{tabular}{|c|c|c|}
 \hline &$\nu SM$ sector &Dark sector\\
 \hline Local gauge groups
 &$SU(2)_L\otimes U(1)_{Y}$ &$SU(2)_D\otimes U(1)_{D'}$\\
 \hline Gauge fields
 &$W_{\mu}(3,1),\; B_{\mu}(1,1)$ &$W_{\mu}^{D}(3,1),\; B_{\mu}^{D}(1,1)$\\
 \hline Chiral fermion fields
 &$l_{L}=\left(\begin{array}{c}\nu_{L}^{0}\\e_{L}^{-}\end{array}\right)(2,-1)_{-1}$, &$l_{L}^{D}=\left(\begin{array}{c}(\chi_{R}^{0})^{c}\\\chi_{L}^{-}\end{array}\right)(2,-1)_{0}$,\\
 &$e_{R}^{-}(1,-2)_{-1},\; N_{R}^{0}(1,0)_{-1}$ &$l_{R}^{D}=\left(\begin{array}{c}(N_{L}^{0})^{c}\\\nu_{R}^{-}\end{array}\right)(2,-1)_{0}$\\
 \hline Scalar fields
 &$H=\left(\begin{array}{c}H^{0}\\H^{-}\end{array}\right)(2,-1)_{0}$ &$\Phi=\left(\begin{array}{c}\Phi^{0}\\\Phi^{-}\end{array}\right)(2,-1)_{-1}$,\\
 & &$\phi^{-}(1,-2)_{0}$\\
 \hline Inflation field & &$\phi^{0}(1,0)_{0}$\\
 \hline Global $B-L$ symmetry &\multicolumn{2}{|c|}{$U(1)_{B-L}$} \\
 \hline Discrete $Z_{4}$ symmetry &\multicolumn{2}{|c|}{$f_{L}\rightarrow i^{n}f_{L},\; f_{R}\rightarrow (-i)^{n}f_{R},\; S\rightarrow (-1)^{n}S,\; n=1,2,3$} \\
 &\multicolumn{2}{|c|}{$f_{L}=(l_{L},l_{L}^{D}),\; f_{R}=(e_{R},N_{R},l_{R}^{D}),\; S=(H,\Phi,\phi^{-},\phi^{0})$}\\ \hline
\end{tabular}
 \caption{The model particle contents and its symmetries. The gauge quantum number of each field is inside its right side bracket and its $B-L$ number is the right subscript of the bracket. The right superscript number of each dark component is its dark electric charge $Q^{D}=I_{3}^{D}+\frac{D'}{2}$ similar to the $\nu SM$ particle electric charge $Q=I_{3}^{L}+\frac{Y}{2}$\,. Note that $(\chi_{R}^{0})^{c}=C\overline{\chi_{R}^{0}}^{T}$ and $(N_{L}^{0})^{c}=C\overline{N_{L}^{0}}^{T}$ are chiral antiparticle states, where $C$ is the charge conjugation matrix. Eventually, $N_{L}$ and $N_{R}$ will be combined into a superheavy Dirac fermion, $\nu_{L}$ and $\nu_{R}$ will form a light Dirac neutrino mass, $\chi_{L}$ and $\chi_{R}$ will form a massive Dirac fermion which becomes the CDM. At the low energy, the $\nu SM$ is actually effective to the SM.}
\end{table}
  In Tab. 1, all the $\nu SM$ particles are singlets under the dark gauge groups $G_{Dark}$, likewise, all the dark particles are singlets under the SM groups $G_{SM}$. The global $U(1)_{B-L}$ symmetry, namely the difference between the baryon number and the lepton one is conserved, is held in common in these two sectors, it can therefore connect the $\nu SM$ sector with the dark one. In fact, only the dark doublet scalar $\Phi$ has ``$-1$" unit of $B-L$ number, while the rest of the dark particles have no $B-L$ numbers. The $\nu SM$ particle electric charge is given by $Q=I_{3}^{L}+\frac{Y}{2}$ after the electroweak breaking, similarly, the dark electric charge of each dark component is given by $Q^{D}=I_{3}^{D}+\frac{D'}{2}$ after $\langle\Phi^{0}\rangle\neq0$ breaking the dark gauge symmetry, see the following Eq. (6). $\phi^{-}$ is a dark complex scalar field, its $Q^{D}$ charge is ``$-1$". The dark neutral singlet $\phi^{0}$ without any charges is a real scalar field, it can act as the inflaton in the universe inflation and reheating, refer to \cite{17}.

  All kinds of the chiral fermions in Tab. 1 are Dirac-type without Majorana-type, and they have three generations as usual. $\chi_{R}^{0}$ and $N_{L}^{0}$ have ``$0$" unit of $Q^{D}$ charge, while $\chi_{L}^{-}$ and $\nu_{R}^{-}$ have ``$-1$" unit of $Q^{D}$ charge. $(\chi_{R}^{0})^{c}=C\overline{\chi_{R}^{0}}^{T}$ is a left-handed antiparticle state, while $(N_{L}^{0})^{c}=C\overline{N_{L}^{0}}^{T}$ is a right-handed antiparticle state. Note that $(l_{L}^{D})^{c}=i\tau_{2}C\overline{\l_{L}^{D}}^{T}=((\chi_{L}^{-})^{c},-\chi_{R}^{0})^{T}$ and $(l_{R}^{D})^{c}=i\tau_{2}C\overline{\l_{R}^{D}}^{T}=((\nu_{R}^{-})^{c},-N_{L}^{0})^{T}$ where $\tau_{2}$ is the second Pauli matrix. At the GUT scale, the dark gauge symmetry is broken by $\langle\Phi\rangle$. $N_{R}^{0}$ in the $\nu SM$ sector and $N_{L}^{0}$ in the dark sector will be combined into a superheavy Dirac fermion $N$. At the low energy, the $Q^{D}$ charge is violated by $\langle\phi^{-}\rangle$. $\nu_{L}^{0}$ in the $\nu SM$ sector and $\nu_{R}^{-}$ in the dark sector will be combined into a light Dirac neutrino $\nu$. $\chi_{L}^{-}$ and $\chi_{R}^{0}$ which are in common in the dark sector will be combined into a massive Dirac fermion $\chi$, which will become the CDM in the model. In virtue of the fermion assignments of Tab. 1, it is easily verified that all of the chiral anomalies are completely cancelled in the model, namely, the model is anomaly-free.

  Finally, the model has also a discrete $Z_{4}$ symmetry which is defined in Tab. 1. All the left-handed fermions, all the right-handed fermions, and all the scalars are respectively in the three representations of $Z_{4}$. Note that $l_{L}^{D}$ and $l_{R}^{D}$ have the same quantum numbers but their representations under $Z_{4}$ are different. The $Z_{4}$ symmetry constrains that the chiral fermions have to couple to the scalars, therefore any explicit mass terms of the fermions are prohibited whether they are Dirac-type or Majorana-type.

  We now write the invariant Lagrangian of the model under the above-mentioned symmetries, which is composed of the gauge kinetic energy terms, the Yukawa couplings and the scalar potentials. The dark gauge kinetic energy terms are
\begin{alignat}{1}
 \mathscr{L}_{Dark\:gauge}=
 &\:\mathscr{L}_{pure\:gauge}+\overline{l_{L}^{D}}i\gamma^{\mu}D_{\mu}l_{L}^{D}
  +\overline{l_{R}^{D}}i\gamma^{\mu}D_{\mu}l_{R}^{D}\nonumber\\
 &+(D^{\mu}\Phi)^{\dagger}D_{\mu}\Phi+(D^{\mu}\phi^{-})^{\dagger}D_{\mu}\phi^{-}
  +\frac{1}{2}\partial^{\mu}\phi^{0}\partial_{\mu}\phi^{0}\,,\nonumber\\
 D_{\mu}=&\:\partial_{\mu}+ig_{D}W_{\mu}^{Di}\frac{\tau^{i}}{2}+ig'_{D}B_{\mu}^{D}\frac{D'}{2}\,,
\end{alignat}
  where $g_{D}$ and $g'_{D}$ are two gauge coupling coefficients associated with $SU(2)_{D}\otimes U(1)_{D'}$. $\tau^{i}$ are the three Pauli matrices and $D'$ is the charge operator of $U(1)_{D'}$. Note that $\phi^{0}$ is a real scalar field without any charges, it plays a role of the inflaton.

  The model Yukawa couplings are
\begin{alignat}{1}
 \mathscr{L}_{Y}=
 &\:\overline{l_{L}}Y_{e}e_{R}^{-}i\tau_{2}H^{*}+\overline{l_{L}}Y_{1}N_{R}^{0}H+l_{R}^{DT}CY_{N}N_{R}^{0}\Phi^{*}\nonumber\\
 &+\frac{1}{2}\,l_{R}^{DT}CY_{2}i\tau_{2}l_{R}^{D}\phi^{+}+\frac{1}{2}\,l_{L}^{DT}CY_{\chi}i\tau_{2}l_{L}^{D}\phi^{+}
  +\overline{l_{R}^{D}}Y_{3}l_{L}^{D}\phi^{0}+h.c.\,,
\end{alignat}
  where $C$ is the charge conjugation matrix and $i\tau_{2}=\epsilon_{\alpha\beta}$ is the two-order antisymmetric tensor. The coupling parameters $Y_{e}, Y_{1}$, etc., are all $3\times3$ complex matrices in the flavour space, moreover, the leading matrix element of each coupling matrix should naturally be $\sim \mathcal{O}(1)$. In particular, $Y_{2}$ and $Y_{\chi}$ must be two antisymmetric matrices due to the spinor anticommutativity and the $\tau_{2}$ antisymmetry. In Eq. (2), the $B-L$ symmetry constrains the $Y_{N}$ coupling term, by which the $\nu SM$ sector is connected with the dark one. These Yukawa couplings will give rise to all kinds of the fermion masses after the relevant scalar fields developing their non-vanishing vacuum expectation values. In addition, the $\phi^{0}$ inflaton decay will lead to the matter-antimatter asymmetry. The $Z_{4}$ symmetry prohibits the explicit mass term of $\overline{l_{L}^{D}}M l_{R}^{D}$ even if it satisfies all of the gauge symmetries, so the $\chi$ fermion can not mix with the neutrino $\nu$. This can guarantee the $\chi$ stability so that $\chi$ will eventually become the CDM. In a word, Eq. (2) plays key roles in the origins of the neutrino mass, the matter-antimatter asymmetry and the CDM.

  The full scalar potentials are
\begin{alignat}{1}
 V_{S}=&\:\mu^{2}_{H}H^{\dagger}H+\mu^{2}_{\Phi}\Phi^{\dagger}\Phi
  +\mu^{2}_{\phi^{-}}\phi^{+}\phi^{-}+\frac{1}{2}\mu^{2}_{\phi^{0}}\phi^{02}\nonumber\\
 &+\lambda_{H}(H^{\dagger}H)^{2}+\lambda_{\Phi}(\Phi^{\dagger}\Phi)^{2}
  +\lambda_{\phi^{-}}(\phi^{+}\phi^{-})^{2}+\frac{\lambda_{\phi^{0}}}{4}\phi^{04}\nonumber\\
 &+2\lambda_{1}H^{\dagger}H\Phi^{\dagger}\Phi+2\lambda_{2}H^{\dagger}H\phi^{+}\phi^{-}
  +2\lambda_{3}\Phi^{\dagger}\Phi\phi^{+}\phi^{-}\nonumber\\
 &+(\lambda_{4}H^{\dagger}H+\lambda_{5}\Phi^{\dagger}\Phi+\lambda_{6}\phi^{+}\phi^{-})\phi^{02}\,.
\end{alignat}
  Furthermore, we assume that all kinds of the parameters in Eq. (3) are constrained as follows,
\begin{alignat}{1}
 &[\lambda_{H},\lambda_{\Phi},\lambda_{\phi_{\phi^{-}}},\lambda_{\phi^{0}}]\sim0.1>0\,,\hspace{0.3cm}
  10^{-6}\lesssim[\lambda_{1},\lambda_{2},\cdots,\lambda_{6}]\lesssim 10^{-2}\,,\nonumber\\
 &\mu^{2}_{\Phi}\approx-\frac{v_{\Phi}^{2}}{\lambda_{\Phi}}\sim-\Lambda^{2}_{GUT},\hspace{0.3cm}\mu^{2}_{H}<-\lambda_{1}v_{\Phi}^{2}\,,\hspace{0.3cm}
  \mu^{2}_{\phi^{-}}<-\lambda_{3}v_{\Phi}^{2}\,,\hspace{0.3cm}\mu^{2}_{\phi^{0}}>-\lambda_{5}v_{\Phi}^{2}\,,
\end{alignat}
  where $\Lambda_{GUT}\approx10^{16}$ GeV is the energy scale of the grand unification and $v_{\Phi}$ is the vacuum expectation value of $\Phi$, see the following Eq. (5). In Eq. (3), it is very believable that the self-interaction of each scalar field is stronger but the interactions among them are weaker, so those interactive coupling parameters are naturally much smaller than those self-coupling parameters in Eq. (4). In addition, those limits of $\mu^{2}_{\Phi},\mu^{2}_{H},\mu^{2}_{\phi^{-}},\mu^{2}_{\phi^{0}}$ are necessary in order to accomplish the spontaneous breakings of the model symmetries. In brief, the limits of Eq. (4) can lead that at the $\Lambda_{GUT}$ scale $\Phi$ earlier develops a non-vanishing vacuum expectation value, later $H$ and $\phi^{-}$ are respectively induced to develop non-vanishing vacuum expectation values at low-energy scale, but $\phi^{0}$ always keeps a vanishing vacuum expectation value, as a result, the symmetry breakings proceed along the chain of the following Eq. (6).

  Based on the limits of Eq. (4), we can derive the vacuum configurations from the $V_{S}$ minimum. The vacua of $H$ and $\Phi$ are necessarily along the directions of their neutral components. The detailed results are as follows,
\begin{alignat}{1}
 &\langle H\rangle=\frac{v_{H}}{\sqrt{2}}
  \left(\begin{array}{c}1\\0\end{array}\right),\hspace{0.3cm} \langle\Phi\rangle=\frac{v_{\Phi}}{\sqrt{2}}\left(\begin{array}{c}1\\0\end{array}\right),\hspace{0.3cm} \langle\phi^{-}\rangle=\frac{v_{\phi}}{\sqrt{2}}\,,\hspace{0.3cm} \langle\phi^{0}\rangle=0\,,\nonumber\\
 &\left(\begin{array}{c}v_{H}^{2}\\v_{\Phi}^{2}\\v_{\phi}^{2}\end{array}\right)=
  \left(\begin{array}{ccc}\lambda_{H}&\lambda_{1}&\lambda_{2}\\\lambda_{1}&\lambda_{\Phi}&\lambda_{3}\\
  \lambda_{2}&\lambda_{3}&\lambda_{\phi^{-}}\end{array}\right)^{-1}
  \left(\begin{array}{c}-\mu^{2}_{H}\\-\mu^{2}_{\Phi}\\-\mu^{2}_{\phi^{-}}\end{array}\right),\nonumber\\
 &v_{\phi}\sim10\:\mathrm{GeV}<v_{H}\approx246\:\mathrm{GeV}\ll
  v_{\Phi}\sim10^{16}\:\mathrm{GeV},
\end{alignat}
  where $v_{\Phi}$ is reasonably about the $\Lambda_{GUT}$ scale, $v_{H}$ is namely the electroweak breaking scale which has been fixed by the SM physics, and $v_{\phi}$ is actually the dark electric charge violating scale which will be determined by the dark sector physics. The vacuum stability mathematically requires that all the ordered principal minors of $Det[\lambda_{ij}]$ must be positive, where $Det[\lambda_{ij}]$ is the determinant corresponding to the $3\times3$ matrix $(\lambda_{ij})$ in Eq. (5), this is also equivalent to $M^{2}_{S}$ in the following Eq. (7) being positive definite. It is easily verified that the limits of Eq. (4) can sufficiently satisfy this condition. In short, the limits of Eq. (4) are natural and reasonable, they can ensure both the vacuum stability, namely Eq. (5), and the model symmetry breaking chain, namely Eq. (6).

  According to the assignments of Tab. 1 and the relations in Eq. (5), the model symmetries are spontaneously broken step by step through the following breaking chain,
\begin{alignat}{1}
 &U(1)_{B-L}^{global}\otimes SU(2)_{D}\otimes U(1)_{D'}
  \xrightarrow{\langle\Phi\rangle} U(1)_{B-L-L^{D}}^{global}\otimes U(1)_{Q^{D}}^{local}
  \xrightarrow{\langle\phi^{-}\rangle} U(1)_{B-L-L^{D}}^{global}\,,\nonumber\\
 &SU(2)_{L}\otimes U(1)_{Y}\xrightarrow{\langle H\rangle} U(1)_{Q}^{em}\,,\nonumber\\
 &Z_{4}\xrightarrow{\langle\Phi\rangle,\langle H\rangle,\langle\phi^{-}\rangle} nothing,\nonumber\\
 &L^{D}=-2I_{3}^{D}\,,\hspace{0.3cm} B-L-L^{D}=B-L+2I_{3}^{D}\,,\hspace{0.3cm}
  Q^{D}=I_{3}^{D}+\frac{D'}{2}\,,\hspace{0.3cm} Q=I_{3}^{L}+\frac{Y}{2}\,,
\end{alignat}
  where I define $L^{D}$ as the dark lepton number which is derived from $I^{D}_{3}$ of $SU(2)_{D}$, and $Q^{D}$ is the dark electric charge similar to the $\nu SM$ electric charge $Q$. The $Q^{D}$ charge of each dark component has been given in Tab. 1. It is easily verified that all the dark chiral fermions, namely $\chi_{L}^{-},\chi_{R}^{0},N_{L}^{0},\nu_{R}^{-}$, have $L^{D}=1$, while all the dark chiral anti-fermions, namely $\chi_{L}^{c},\chi_{R}^{c},N_{L}^{c},\nu_{R}^{c}$, has $L^{D}=-1$. At the first step of the dark symmetry breaking, $\langle\Phi\rangle\sim 10^{16}$ GeV breaks both the global $U(1)_{B-L}$ and the local $SU(2)_{D}\otimes U(1)_{D'}$, but the global $U(1)_{B-L-L^{D}}$ and the local $U(1)_{Q^{D}}$ are conserved as two residual symmetries because the up-type component $\Phi^{0}$ has both $B-L-L^{D}=0$ and $Q^{D}=0$. This breaking is very analogous to the later electroweak breaking. As a result, some dark particles, for example, $\Phi^{0},W_{\mu}^{D},N$, are generated superheavy masses around the $v_{\Phi}$ scale through the Higgs mechanism. At the second step of the dark symmetry breaking, $\langle\phi^{-}\rangle\sim 10$ GeV further breaks the local $U(1)_{Q^{D}}$ but still keeps the global $B-L-L^{D}$ conservation because $\phi^{-}$ has $Q^{D}=-1$ and $B-L-L^{D}=0$. As a result, the dark neutral scalar $\rho^{0}$ as the successor of $\phi^{-}$ (see Eq. (7)), the dark photon $A_{\mu}^{D}$ associated with $U(1)_{Q^{D}}$ (see Eq. (8)), and the CDM fermion $\chi$ obtain their masses around the $v_{\phi}$ scale. In the $\nu SM$ sector, $\langle H\rangle\sim 10^{2}$ GeV accomplishes the electroweak breaking and gives rise to the SM particle masses around the $v_{H}$ scale. The gauge symmetry breakings is actually accompanied with the $Z_{4}$ symmetry breaking. Eventually, the residual symmetries are the local $Q$ conservation in the $\nu SM$ sector and the global $B-L-L^{D}$ conservation in common in these two sectors.

  After the above-mentioned symmetry breakings are completed, all kinds of particle masses and mixings are generated through the Higgs mechanism. In the scalar sector, there are now three neutral bosons $h^{0},\Phi_{Re}^{0},\rho^{0}$ besides the inflaton $\phi^{0}$, their masses and mixing are given by the following relations,
\begin{alignat}{1}
 &H\rightarrow\frac{h^{0}+v_{H}}{\sqrt{2}}\left(\begin{array}{c}1\\0\end{array}\right),\hspace{0.3cm}
  \Phi\rightarrow\frac{\Phi^{0}_{Re}+v_{\Phi}}{\sqrt{2}}\left(\begin{array}{c}1\\0\end{array}\right),\hspace{0.3cm} \phi^{-}\rightarrow\frac{\rho^{0}+v_{\phi}}{\sqrt{2}}\,,\nonumber\\
 &M^{2}_{S}=2\left(\begin{array}{ccc}\lambda_{H}v_{H}^{2}&\lambda_{1}v_{H}v_{\Phi}&\lambda_{2}v_{H}v_{\phi}\\
  \lambda_{1}v_{H}v_{\Phi}&\lambda_{\Phi}v_{\Phi}^{2}&\lambda_{3}v_{\Phi}v_{\phi}\\
  \lambda_{2}v_{H}v_{\phi}&\lambda_{3}v_{\Phi}v_{\phi}&\lambda_{\phi^{-}}v_{\phi}^{2}\end{array}\right) \xrightarrow{diagonalizing}\left(\begin{array}{ccc}M^{2}_{h^{0}}&0&0\\
  0&M^{2}_{\Phi^{0}}&0\\0&0&M^{2}_{\rho^{0}}\end{array}\right), \nonumber\\
 &M_{h^{0}}\approx\sqrt{2\lambda_{H}}\,v_{H}\,,\hspace{0.3cm}
  M_{\Phi^{0}}\approx\sqrt{2\lambda_{\Phi}}\,v_{\Phi}\,,\hspace{0.3cm} M_{\rho^{0}}\approx\sqrt{2\lambda_{\phi^{-}}}\,v_{\phi}\,,\hspace{0.3cm} M_{\phi^{0}}\approx\sqrt{\mu^{2}_{\phi^{0}}+\lambda_{5}v_{\Phi}^{2}}\,.
\end{alignat}
  Here the four dark scalar components $\Phi^{\pm},\Phi^{0}_{Im},\phi^{-}_{Im}$ have been transferred into the dark gauge sector to yield the masses of the four dark gauge fields $W_{\mu}^{D\pm},Z_{\mu}^{D0},A_{\mu}^{D0}$ through the Higgs mechanism, see the following Eq. (8). In view of the parameter values in Eqs. (4) and (5), the mixing angles of $M^{2}_{S}$ are all smaller, for example, the mixing angle between $h^{0}$ and $\rho^{0}$ is $\sim\frac{\lambda_{2}v_{H}v_{\phi}}{\lambda_{H}v_{H}^{2}-\lambda_{\phi^{-}}v_{\phi}^{2}}<10^{-2}$\,, so the three eigenvalues of $M^{2}_{S}$ is approximately equal to its diagonal elements. $h^{0}$ is exactly the SM Higgs boson with $M_{h^{0}}\approx125$ GeV. $M_{\Phi^{0}}$ is close to $\Lambda_{GUT}$\,, therefore, $\Phi^{0}_{Re}$ can not appear in the low-energy phenomena. $M_{\rho^{0}}$ is about several GeVs, so $\rho^{0}$ is a relatively light dark scalar. Explicitly, the inflaton $\phi^{0}$ has no mixing with the other neutral bosons, $M_{\phi^{0}}$ is derived from the two contributions which include the original mass $\mu_{\phi^{0}}$ and the induced mass from $\langle\Phi\rangle$, however, its reasonable value should be $M_{\phi^{0}}\sim10^{11}$ GeV which is close to the universe reheating temperature of $\sim10^{12}$ GeV \cite{18}.

  In the dark gauge sector, the masses and mixing of the dark gauge fields are very similar to the case of the SM weak gauge fields. The only difference between these two cases is that the dark electric charge $U(1)_{Q^{D}}$ is eventually violated by $\langle\phi^{-}\rangle$ so that the dark photon is massive, whereas the $\nu SM$ electric charge $U(1)_{Q}^{em}$ is always conserved so that the photon is massless. We can derive from Eq. (1) the following relations,
\begin{alignat}{1}
 &\mathscr{L}_{gauge\:mass}=\frac{g_{D}^{2}v_{\Phi}^{2}}{8}(W^{D1}_{\mu}W^{\mu D1}+W^{D2}_{\mu}W^{\mu D2})\nonumber\\
 &\hspace{2.4cm}+\frac{1}{8}(W^{D3}_{\mu},B^{D}_{\mu})\left(\begin{array}{cc}g_{D}^{2}v_{\Phi}^{2}&-g_{D}g'_{D}v_{\Phi}^{2}\\
  -g_{D}g'_{D}v_{\Phi}^{2}&g'^{2}_{D}(v_{\Phi}^{2}+4v_{\phi}^{2})\end{array}\right)
  \left(\begin{array}{c}W^{\mu D3}\\B^{\mu D}\end{array}\right)\nonumber\\
 &\xrightarrow{diagonalizing}M^{2}_{W^{D}}W^{D+}_{\mu}W^{\mu D-}+\frac{1}{2}(Z^{D}_{\mu},A^{D}_{\mu})
  \left(\begin{array}{cc}M^{2}_{Z^{D}}&0\\0&M^{2}_{A^{D}}\end{array}\right)
  \left(\begin{array}{c}Z^{\mu D}\\A^{\mu D}\end{array}\right),\nonumber\\
 &M_{W^{D}}=\frac{g_{D}v_{\Phi}}{2}\,,\hspace{0.3cm}
  M_{Z^{D}}=\frac{M_{W^{D}}}{cos\theta_{D}}\,,\hspace{0.3cm} M_{A^{D}}=e_{D}v_{\phi}\,,\nonumber\\
 &W_{\mu}^{D\mp}=\frac{W_{\mu}^{D1}\pm iW_{\mu}^{D2}}{\sqrt{2}}\,,\hspace{0.2cm}
  Z_{\mu}^{D}=cos\theta_{D}W_{\mu}^{D3}-sin\theta_{D}B_{\mu}^{D},\hspace{0.2cm} A_{\mu}^{D}=sin\theta_{D}W_{\mu}^{D3}+cos\theta_{D}B_{\mu}^{D},\nonumber\\
 &sin\theta_{D}=\frac{g'_{D}}{\sqrt{g^{2}_{D}+g'^{2}_{D}}}\,,\hspace{0.3cm} e_{D}=g_{D}sin\theta_{D},\hspace{0.3cm}
  \widetilde{Q}^{D}=\frac{I_{3}^{D}-sin^{2}\theta_{D}Q^{D}}{cos\theta_{D}}\,,\nonumber\\
 &D_{\mu}\rightarrow\partial_{\mu}+i\frac{g_{D}}{\sqrt{2}}(W_{\mu}^{D+}\tau^{+}+W_{\mu}^{D-}\tau^{-})
  +ig_{D}Z_{\mu}^{D}\widetilde{Q}^{D}+ie_{D}A_{\mu}^{D}Q^{D}\,,\nonumber\\
 &\mathscr{L}_{Dark\:gauge}\xrightarrow{including}e_{D}A_{\mu}^{D}
  (\overline{\chi_{L}^{-}}\gamma^{\mu}\chi_{L}^{-}+\overline{\nu_{R}^{-}}\gamma^{\mu}\nu_{R}^{-})
  +\frac{e_{D}^{2}}{2}A_{\mu}^{D}A^{\mu D}(\rho^{0}+v_{\phi})^{2}\,,
\end{alignat}
  where all kinds of notations are self-explanatory. The dark $sin\theta_{D},e_{D},\widetilde{Q}^{D}$ and their sizes are all analogous to ones of the $\nu SM$. $W_{\mu}^{D\mp}$ and $Z_{\mu}^{D}$ are all superheavy, but $A_{\mu}^{D}$ is light, $M_{A^{D}}$ is about several GeVs. However, it should be emphasized that below the $v_{\phi}$ scale, the dark electric charge is violated and its physical meaning is thereupon vanishing, therefore, all of the dark charged states $W_{\mu}^{D\mp},\chi_{L}^{-},\nu_{R}^{-}$ should eventually be regarded as the neutral states, namely, the ``$\mp$" superscripts can be disregarded or picked off.

  In the Yukawa sector, the Yukawa couplings undergo evolutions as follows,
\begin{alignat}{1}
 \mathscr{L}_{Y}\xrightarrow{\langle\Phi\rangle}
 &\:\overline{l_{L}}Y_{e}e_{R}^{-}i\tau_{2}H^{*}+\overline{l_{L}}Y_{1}N_{R}^{0}H
  -\overline{N_{L}^{0}}M_{N}N_{R}^{0}+\overline{N_{L}^{0}}Y_{2}\nu_{R}^{-}\phi^{+}\nonumber\\
 &+\overline{\chi_{R}^{0}}Y_{\chi}\chi_{L}^{-}\phi^{+}
  +\overline{N_{L}^{0}}Y_{3}^{*}\chi_{R}^{0}\phi^{0}
  +\overline{\nu_{R}^{-}}Y_{3}\chi_{L}^{-}\phi^{0}+h.c.\,,\nonumber\\
 \xrightarrow{effective}&\:\overline{l_{L}}Y_{e}e_{R}^{-}i\tau_{2}H^{*}
  +\overline{l_{L}}HY_{1}M_{N}^{-1}Y_{2}\nu_{R}^{-}\phi^{+}
  +\overline{\chi_{R}^{0}}Y_{\chi}\chi_{L}^{-}\phi^{+}\nonumber\\
 &+\overline{l_{L}}HY_{1}M_{N}^{-1}Y_{3}^{*}\chi_{R}^{0}\phi^{0}
  +\overline{\nu_{R}^{-}}Y_{3}\chi_{L}^{-}\phi^{0}+h.c.\,,\nonumber\\
 \xrightarrow{\langle H\rangle,\langle\phi^{-}\rangle}
 &-\overline{e_{L}^{-}}M_{e}e_{R}^{-}-\overline{\nu_{L}}M_{\nu}\nu_{R}
  -\overline{\chi_{R}}M_{\chi}\chi_{L}\nonumber\\
 &+\overline{\nu_{L}}Y_{1}\frac{v_{H}}{\sqrt{2}M_{N}}Y_{3}^{*}\chi_{R}\phi^{0}
  +\overline{\nu_{R}}Y_{3}\chi_{L}\phi^{0}+h.c.\,,\nonumber\\
 M_{N}=&-\frac{v_{\Phi}}{\sqrt{2}}Y_{N}\,,\hspace{0.3cm}
  M_{e}=\frac{v_{H}}{\sqrt{2}}Y_{e}\,,\hspace{0.3cm}
  M_{\chi}=-\frac{v_{\phi}}{\sqrt{2}}Y_{\chi}\,,\nonumber\\
 M_{\nu}=&-Y_{1}\frac{v_{H}v_{\phi}}{2M_{N}}Y_{2}=
  -\frac{v_{H}v_{\phi}}{2m_{N_{1}}}Y_{1}\frac{m_{N_{1}}}{M_{N}}Y_{2}\,,
\end{alignat}
  where $m_{N_{1}}$ is the lightest one of the three mass eigenvalues of the mass matrix $M_{N}$ such as $(m_{N_{1}},m_{N_{2}},m_{N_{3}})$, thus the dimensionless matrix $\frac{m_{N_{1}}}{M_{N}}$ is certainly $\sim \mathcal{O}(1)$. In Eq. (9), at first $\langle\Phi\rangle$ at the GUT scale breaks the dark gauge symmetry, and yields the superheavy Dirac mass $M_{N}$ through the combination of $N_{L}^{0}$ and $N_{R}^{0}$, accordingly, the doublet couplings related to $l_{L}^{D}$ and $l_{R}^{D}$ are decomposed into their component couplings. Obviously, both the $B-L-L^{D}$ number and the $Q^{D}$ charge are conversed, furthermore, $N$ actually acts as a mediator between the dark sector and the $\nu SM$ one. At the second step, the universe temperature drops to the $M_{\phi^{0}}$ scale, $N$ has been decoupling and it is integrated out because $M_{N}$ is several orders of magnitude heavier than $M_{\phi^{0}}$, thus we obtain an effective Yukawa couplings with the 5-dimensional operators suppressed by $M_{N}^{-1}$. The inflaton $\phi^{0}$ plays a leading actor at this stage. The $\phi^{0}$ decay will not only make the universe inflation and reheating, but also lead to the matter-antimatter asymmetry. At the last step, $\langle H\rangle$ and $\langle\phi^{-}\rangle$ at the low-energy scale break the electroweak symmetry and the $Q^{D}$ charge, respectively, and they generate the Dirac masses of the charged lepton, the neutrino $\nu$, and the CDM $\chi$. Note that $\chi_{L},\nu_{R}$ whose ``$-$" superscripts are picked off now become the neutral states since the dark electric charge has now been meaningless or vanishing. $M_{e}$ is well known around one GeV. $M_{\chi}$ is about several GeVs, $\chi$ will become the CDM in the next section discussion. By contrast, $M_{\nu}$ is only Sub-eV due to the $M_{N}^{-1}$ suppression. If $m_{N_{1}}\sim10^{13}$ GeV, we can then obtain $M_{\nu}\sim10^{-2}$ eV which is exactly fitting with the experimental data. The Feynman diagram of generating neutrino mass is shown in Fig. 1, this mechanism is a Dirac-type seesaw, obviously, it is different from the usual Majorana-type seesaw \cite{6}.
\begin{figure}
 \centering
 \includegraphics[totalheight=5cm]{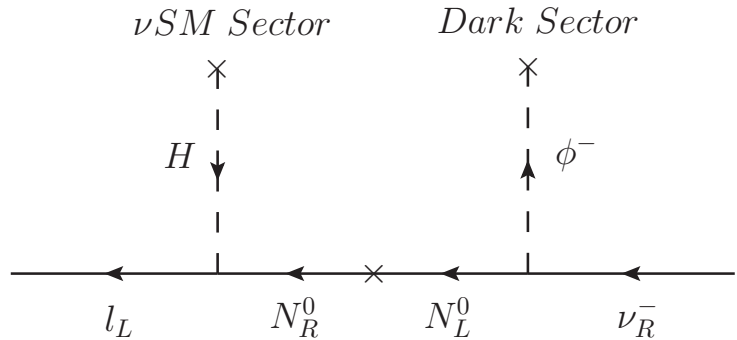}
 \caption{The diagram of generating Dirac neutrino mass for $\nu_{L}$ and $\nu_{R}$. The superheavy Dirac mass of $N^{0}_{L}$ and $N^{0}_{R}$ is generated by $\langle\Phi\rangle$ at the GUT scale, which breaks $SU(2)_D\otimes U(1)_{D'}$ down to $U(1)_{Q^{D}}$ and leads to the new $B-L-L^{D}$ conservation instead of the $B-L$ symmetry. At the low energy $N$ is integrated out, $\langle\phi^{-}\rangle$ further violates the ${Q^{D}}$ charge so that $\nu_{R}$ eventually becomes a neutral state. Note that $\triangle(B-L)=\triangle L^{D}=1$ but $\triangle(B-L-L^{D})=0$ in the diagram. }
\end{figure}

  In the first two rows in Eq. (9), the three matrices of $Y_{e},M_{N},Y_{\chi}$ can simultaneously be diagonalized by the relevant left-handed and right-handed flavour basis rotations, thus the three matrices of $Y_{1},Y_{2},Y_{3}$ are all non-diagonal ones, they certainly contain some irremovable complex phases, so there are new $CP$-violating sources besides the CKM phase in the quark sector. Under the mass eigenstate bases of $M_{e},M_{N},M_{\chi}$, the non-diagonal $M_{\nu}$ contains the full information of the neutrino masses and mixing. In particular it should be pointed out that because $Y_{\chi}$ is an antisymmetric matrix, the three eigenvalues of $M_{\chi}$ are such spectrum as ${m_{\chi_{1}}=0<m_{\chi_{2}}=m_{\chi_{3}}}$. Similarly, the eigenvalue spectrum of $M_{\nu}$ has two special modes due to the antisymmetric $Y_{2}$ factor of the $M_{\nu}$ matrix. The rigorous mathematical results are as follows, i) the normal order such as ${m_{\nu_{1}}=0<m_{\nu_{2}}<m_{\nu_{3}}}$ if the three eigenvalues of $M_{N}$ are a hierarchy, ii) the inverted order such as ${m_{\nu_{3}}=0<m_{\nu_{1}}\approx m_{\nu_{2}}}$ if the three eigenvalues of $M_{N}$ are an approximate degeneracy. In the model, the three eigenvalues of $M_{N}$ are actually required to be such hierarchy as the following Eq. (10), therefore the neutrino mass spectrum should be taken as the normal order mode. Based on the discussions in this Section, finally, we summarize that the particle mass spectrum of the model is such relations as
\begin{alignat}{1}
 &(M_{A},m_{\chi_{1}},m_{\nu_{1}})=0<m_{\nu_{2}}< m_{\nu_{3}}\sim 0.05\:\mathrm{eV}\ll M_{e}\sim[10^{-3}-1]\:\mathrm{GeV}\nonumber\\
 <&(m_{\chi_{2}}=m_{\chi_{3}},M_{A^{D}},M_{\rho^{0}})\sim[3-5]\:\mathrm{GeV}
 <(M_{W},M_{Z},M_{h^{0}})\sim100\:\mathrm{GeV}\nonumber\\
 \ll&M_{\phi^{0}}\sim 10^{11}\:\mathrm{GeV}<m_{N_{1}}\sim10^{13}\:\mathrm{GeV}
  <m_{N_{2}}<(m_{N_{3}},M_{W^{D}},M_{Z^{D}},M_{\Phi^{0}})\sim 10^{15}\:\mathrm{GeV},
\end{alignat}
  where $M_{A}$ is the photon mass. In short, the mass relations of Eq. (10) will lead to successful explanations for the matter-antimatter asymmetry and the CDM in the following Sections.

\vspace{0.6cm}
\noindent\textbf{III. Baryon Asymmetry and Asymmetric CDM from Inflaton Decay}

\vspace{0.3cm}
  The model can explain the common origin of the baryon asymmetry and the asymmetric CDM, concretely, these two things together arise from the inflaton $\phi^{0}$ decay in the era of the universe inflation and reheating. The discussions about the inflaton making the inflation and reheating are beyond this paper scope, one can refer to \cite{17,18}, here we are only concerned with its particle physics aspect. From the effective Yukawa couplings in Eq. (9), which conserve both the $B-L-L^{D}$ number and the $Q^{D}$ charge, we can see that the inflaton $\phi^{0}$ has only two decay modes at the tree level, i) the two-body decays of $\phi^{0}\rightarrow \nu_{R}^{-}+\widetilde{\chi_{L}^{-}}$ and $\phi^{0}\rightarrow \widetilde{\nu_{R}^{-}}+\chi_{L}^{-}$, hereinafter a notation with the tilde denotes its $CP$ conjugate state, ii) the three-body decays of $\phi^{0}\rightarrow l_{L}+\widetilde{H}+\widetilde{\chi_{R}^{0}}$ and $\phi^{0}\rightarrow \widetilde{l_{L}}+H+\chi_{R}^{0}$, which are suppressed by $M_{N}^{-1}$. The two-body decays are essentially dedicated to the universe inflation and reheating, while the three-body decays can successfully lead to the baryon asymmetry and the asymmetric CDM.

  Fig. 2 shows the tree and loop diagrams of $\phi^{0}\rightarrow l_{L}+\widetilde{H}+\widetilde{\chi_{R}^{0}}$ on the basis of the effective Yukawa couplings in Eq. (9).
\begin{figure}
 \centering
 \includegraphics[totalheight=5.5cm]{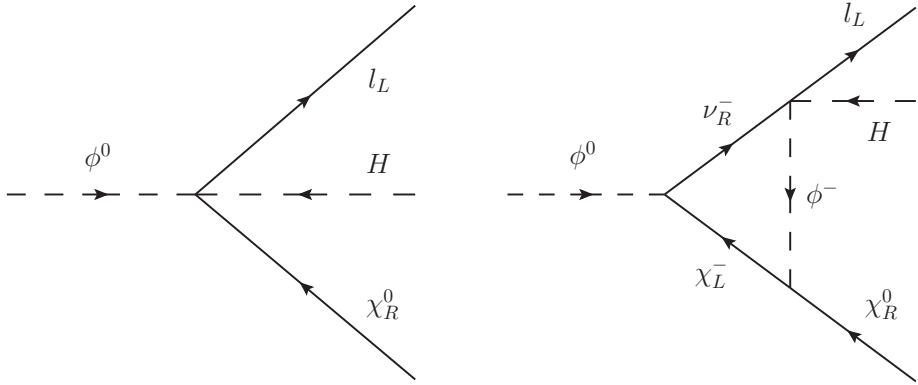}
 \caption{The tree and loop diagrams of the inflaton three-body decay $\phi^{0}\rightarrow l_{L}+\widetilde{H}+\widetilde{\chi_{R}^{0}}$\,. The decay process has $\triangle(B-L)=\triangle L^{D}=-1$ but $\triangle(B-L-L^{D})=0$, so it can simultaneously generate the $B-L$ asymmetry in the $\nu SM$ sector and the $L^{D}$ asymmetry in the dark sector, which eventually lead to the baryon asymmetry and the asymmetric CDM $\chi$\,.}
\end{figure}
  The decay in Fig. 2 has the following characteristics. This decay explicitly violates both ``$-1$" unit of $B-L$ number and ``$-1$" unit of $L^{D}$ number but converses the total $B-L-L^{D}$ number, namely
\ba
 \triangle(B-L)=\triangle L^{D}=-1,\hspace{0.3cm} \triangle(B-L-L^{D})=0.
\ea
  In addition, the decay rate of Fig. 2 is different from one of its $CP$ conjugate process through the interference between the tree diagram and the loop one. The $CP$ asymmetry of these two decay rates is defined and calculated as follows,
\begin{alignat}{1}
 &\varepsilon=\frac{\Gamma[\phi^{0}\rightarrow l_{L}+\widetilde{H}+\widetilde{\chi_{R}^{0}}]
  -\Gamma[\phi^{0}\rightarrow\widetilde{l_{L}}+H+\chi_{R}^{0}]}{\Gamma_{\phi^{0}}}\nonumber\\
 &\hspace{0.2cm}=\frac{(\frac{M_{\phi^{0}}}{m_{N_{1}}})^{2}}{384\pi^{2}}\,
  \frac{Im[Tr[(Y_{1}\frac{m_{N_{1}}}{M_{N}}Y_{2})Y_{3}Y_{\chi}^{\dagger}
  (Y_{1}\frac{m_{N_{1}}}{M_{N}}Y_{3}^{*})^{\dagger}]]}{Tr[Y_{3}Y_{3}^{\dagger}]}\,,\nonumber\\
 &\Gamma_{\phi^{0}}=\Gamma[\phi^{0}\rightarrow l_{L}+\widetilde{H}+\widetilde{\chi_{R}^{0}}]
  +\Gamma[\phi^{0}\rightarrow\nu_{R}^{-}+\widetilde{\chi_{L}^{-}}]+CP\:conjugate\,,\nonumber\\
 &\Gamma[\phi^{0}\rightarrow l_{L}+\widetilde{H}+\widetilde{\chi_{R}^{0}}]=
  \frac{M_{\phi^{0}}(\frac{M_{\phi^{0}}}{m_{N_{1}}})^{2}}{192(2\pi)^{3}}\,
  Tr[(Y_{1}\frac{m_{N_{1}}}{M_{N}}Y_{3}^{*})(Y_{1}\frac{m_{N_{1}}}{M_{N}}Y_{3}^{*})^{\dagger}]\,,\nonumber\\
 &\Gamma[\phi^{0}\rightarrow\nu_{R}^{-}+\widetilde{\chi_{L}^{-}}]=\frac{M_{\phi^{0}}}{16\pi}Tr[Y_{3}Y_{3}^{\dagger}]\,,
\end{alignat}
  where $\Gamma_{\phi^{0}}$ is the total decay width of $\phi^{0}$. Obviously, the three-body decay rate is far smaller than the two-body one on because of the twofold suppressions of the phase space factor and the $(\frac{M_{\phi^{0}}}{m_{N_{1}}})^{2}$ factor, so $\Gamma_{\phi^{0}}$ is essentially dominated by the two-body decay width. In the calculation of $\varepsilon$, the imaginary part of the loop integration factor is totally derived from the two-point function $Im[B_{0}((p_{l}+p_{\widetilde{H}})^{2},M^{2}_{\phi^{-}},M^{2}_{\nu})]$ $=i\pi$. Note that the matrix factor $(Y_{1}\frac{m_{N_{1}}}{M_{N}}Y_{2})$ in the $\varepsilon$ equality, it is actually related to the neutrino mass matrix by $(Y_{1}\frac{m_{N_{1}}}{M_{N}}Y_{2})=-\frac{2m_{N_{1}}}{v_{H}v_{\phi}}M_{\nu}$, see Eq. (9). The $CP$ asymmetry $\varepsilon$ is certainly non-vanishing because there are $CP$-violating complex phases in the matrices $Y_{1},Y_{2},Y_{3}$\,. According to the discussions in Section II, the Yukawa matrices and the dimensionless matrix $\frac{m_{N_{1}}}{M_{N}}$ are all $\sim\mathcal{O}(1)$, therefore those two matrix trace factors in the numerator and denominator of the $\varepsilon$ equality are certainly $\sim\mathcal{O}(1)$, thus we can naturally obtain $\varepsilon\sim 10^{-8}$ for $\frac{M_{\phi^{0}}}{m_{N_{1}}}\sim10^{-2}$, which is a reasonable and suitable value for the leptogenesis.

  When the two-body and three-body decay rates in Eq. (12) are in comparison with the universe Hubble expansion rate, a simple calculation gives
\ba
 \Gamma[\phi^{0}\rightarrow l_{L}+\widetilde{H}+\widetilde{\chi_{R}^{0}}]
 \ll H(M_{\phi^{0}})=\frac{1.66\sqrt{g_{*}}M^{2}_{\phi^{0}}}{M_{Pl}}
 \ll \Gamma[\phi^{0}\rightarrow\nu_{R}^{-}+\widetilde{\chi_{L}^{-}}]\,,
\ea
  where $M_{Pl}=1.22\times10^{19}$ GeV and $g_{*}$ is the effective number of relativistic degrees of freedom. At the temperature of $T=M_{\phi^{0}}$, the relativistic states include all the SM particles and these dark particle states $A_{\mu}^{D},\phi^{\mp},\chi_{L}^{-},\chi_{R}^{0},\nu_{R}^{-}$, so we should take $g_{*}=126.5$ in Eq. (13). Eq. (13) indicates that the $\phi^{0}$ three-body decay is severely out-of-equilibrium but the $\phi^{0}$ two-body decay is still in equilibrium. In Section V, we can numerically calculate out $\Gamma[\phi^{0}\rightarrow l_{L}+\widetilde{H}+\widetilde{\chi_{R}^{0}}]/H\approx0.007$ by use of those parameter values of Eq. (23), this indeed demonstrates the above point.

  Up to now, we have completely demonstrated that the $\phi^{0}$ three-body decay as shown in Fig. 2 indeed satisfies Sakharov's three conditions \cite{19}. As a consequence, this decay can simultaneously generate an asymmetry of the $B-L$ number in the $\nu SM$ sector and one of the $L^{D}$ number in the dark sector, these asymmetries normalized to the entropy are given by the relations as follows \cite{20},
\begin{alignat}{1}
 &Y_{B-L}=\frac{n_{B-L}-\widetilde{n}_{B-L}}{s}=\kappa\frac{\varepsilon\triangle(B-L)}{g_{*}}=
  \kappa\frac{\varepsilon(-1)}{g_{*}}\,,\nonumber\\
 &Y_{L^{D}}=\frac{n_{L^{D}}-\widetilde{n}_{L^{D}}}{s}=\kappa\frac{\varepsilon
  \triangle L^{D}}{g_{*}}=\kappa\frac{\varepsilon(-1)}{g_{*}}\,,\nonumber\\
 &Y_{B-L-L^{D}}=\kappa\frac{\varepsilon\triangle(B-L-L^{D})}{g_{*}}=0,
\end{alignat}
  where $g_{*}=126.5$, $s$ is the total entropy density in the $\nu SM$ and dark sectors, $\kappa$ is a dilution factor. The dilution is only from the two 5-dimensional Yukawa interactions suppressed by $M_{N}^{-1}$ in Eq. (9), but they have been severely out-of-equilibrium at $T\leqslant M_{\phi^{0}}$, so the dilution effect is actually very weak, thus we can take $\kappa\approx1$ in Eq. (14).

  After the $\phi^{0}$ decays are over, the universe inflation and reheating are completed, then the universe comes into the radiation-dominated epoch. The universe temperature soon falls below $M_{\phi^{0}}$ as the universe expansion. At the low energy, all kinds of the superheavy particles have been decoupling, the available connection between the $\nu SM$ sector and the dark one is only through the $\lambda_{2}$ scalar coupling in Eq. (3). However, the $\lambda_{2}$ coupling violates neither the $B-L$ number nor the $L^{D}$ number, therefore, the generated $Y_{B-L}$ asymmetry is conserved in the $\nu SM$ sector, while the generated $Y_{L^{D}}$ asymmetry is conversed in the dark sector, but the total $Y_{B-L-L^{D}}$ asymmetry in these two sectors is always zero.

  Before the universe temperature drops to the electroweak scale of $\sim10^{2}$ GeV, the sphaleron process in the $\nu SM$ sector is smoothly put into effect \cite{21}, it can convert a part of the $Y_{B-L}$ asymmetry into the baryon asymmetry. Similarly, before the dark electric charge is violated, the $Y_{L^{D}}$ asymmetry in the dark sector is redistributed among these dark lepton states $\chi_{L}^{-},\chi_{R}^{0},\nu_{R}^{-}$ through the chemical potential equilibrium, see Appendix A. After that, the baryon asymmetry in the $\nu SM$ sector and the $\chi$ asymmetry in the dark sector are eventually fixed, thus they survive to the present day. In the present universe, the asymmetric $\chi$ has become the CDM, see the discussions in the next Section. Therefore, the baryon asymmetry and the CDM $\chi$ asymmetry are given by
\begin{alignat}{1}
 &Y_{B}=c_{s}Y_{B-L}\,,\hspace{0.3cm}\eta_{B}=\frac{n_{B}-\widetilde{n}_{B}}{n_{\gamma}}
  =\frac{s}{n_{\gamma}}Y_{B}\approx6.1\times10^{-10}\,,\nonumber\\
 &Y_{\chi}=c_{\chi}Y_{L^{D}}\,,\hspace{0.3cm}\eta_{\chi}=\frac{n_{\chi}-\widetilde{n}_{\chi}}{n_{\gamma}}
  =\frac{s}{n_{\gamma}}Y_{\chi}=\frac{79}{36}\eta_{B}\,,\nonumber\\
 &c_{s}=\frac{28}{79}\,,\hspace{0.3cm}c_{\chi}=\frac{7}{9}\,,\hspace{0.3cm}\frac{s}{n_{\gamma}}=8.05\,.
\end{alignat}
  $c_{s}$ is the sphaleron conversion coefficient in the $\nu SM$ sector, similarly, $c_{\chi}$ is the conversion coefficient in the dark sector, its derivation is given by Appendix A. The ratio of the total entropy density to the photon number density is equal to $8.05$ in the model, its derivation is given by Appendix B. By use of Eqs. (12) and (14), we can naturally obtain $\eta_{B}\approx6.1\times10^{-10}$ which is the current value of the baryon asymmetry from multiple experiments \cite{1,22}, in addition, the model predicts $\frac{\eta_{\chi}}{\eta_{B}}=\frac{79}{36}$. In conclusion, the inflaton decay as shown in Fig. 2 is the common origin of $\eta_{B}$ and $\eta_{\chi}$ in Eq. (15).

\vspace{0.6cm}
\noindent\textbf{IV. Dark Sector Physics}

\vspace{0.3cm}
  In the radiation phase of the early universe, the dark particle states $A_{\mu}^{D},\phi^{\mp},\chi_{L}^{-},\chi_{R}^{0},\nu_{R}^{-}$ make up of a dark plasma in thermal equilibrium, and the dark electric charge is conversed. When the universe temperature drops to the $v_{\phi}\sim10$ GeV scale, the dark electric charge is violated by $\langle\phi^{-}\rangle$ so that its physical meaning is vanishing, therefore, all of the dark particles eventually become neutral states without any charge. In addition, $\langle\phi^{-}\rangle$ gives rise to the $A_{\mu}^{D},\rho^{0},\chi_{2,3}$ masses about several GeVs except that $\chi_{1}$ is massless. The dark $\nu_{R}$ is combined with the $\nu_{L}$ in $\nu SM$ to yield a tiny Dirac neutrino mass through the Dirac-type seesaw mechanism. As the universe temperature falling, $A_{\mu}^{D},\rho^{0},\chi_{2,3}$ will eventually become non-relativistic particles, while $\chi_{1},\nu_{R},\nu_{L}$ are still relativistic.

  At the low energy, the only available connection between the dark sector and the $\nu SM$ one is the $\lambda_{2}$ scalar coupling in Eq. (3), by which the $\nu SM$ sector can communicate with the dark one. The (a) diagram in Fig. 3 shows the annihilating process $h^{0}+h^{0}\rightarrow \rho^{0}+\rho^{0}$. When the annihilating rate becomes smaller than the universe expansion rate, then the dark sector will decouple from the $\nu SM$ sector, thus these two sectors will be isolated from each other. The decoupling temperature is derived from the following relations,
\begin{alignat}{1}
 &\Gamma[h^{0}+h^{0}\rightarrow \rho^{0}+\rho^{0}]=
  \langle\sigma v_{r}\rangle n_{h^{0}}(T_{D})\lesssim H(T_{D})\,,\nonumber\\
 &\langle\sigma v_{r}\rangle=\frac{\lambda_{2}^{2}}{8\pi M^{2}_{h^{0}}}(1-\frac{3}{2x})\,,\hspace{0.3cm}
  n_{h^{0}}=T_{D}^{3}\left(\frac{x}{2\pi}\right)^{\frac{3}{2}}e^{-x}\,,\hspace{0.3cm}
  x=\frac{M_{h^{0}}}{T_{D}}\,,\nonumber\\
 \Longrightarrow &x\approx
  40.7+ln[M_{h^{0}}\langle\sigma v_{r}\rangle\sqrt{\frac{x}{g_{*}(T_{D})}}\,]\,,\nonumber\\
 \Longrightarrow &T_{D}\sim(7-15)\:\mathrm{GeV}\;\mbox{for}\:10^{-5}\lesssim\lambda_{2}\lesssim10^{-3}\,,
\end{alignat}
  where $v_{r}$ is a relative velocity of two annihilating particles and we have used the thermal average $\langle v_{r}^{2}\rangle=\frac{6}{x}$\,. The total effective number of relativistic degrees of freedom in these two sectors is $g_{*}(T_{D})=70.5$, which is obtained by the third and fourth lines in the following Eq. (21). Eq. (16) indicates that $T_{D}$ is around $v_{\phi}$ by coincidence, in other words, the dark sector is decoupling from the $\nu SM$ sector around the time when the dark electric charge is violated.

\begin{figure}
 \centering
 \includegraphics[totalheight=8cm]{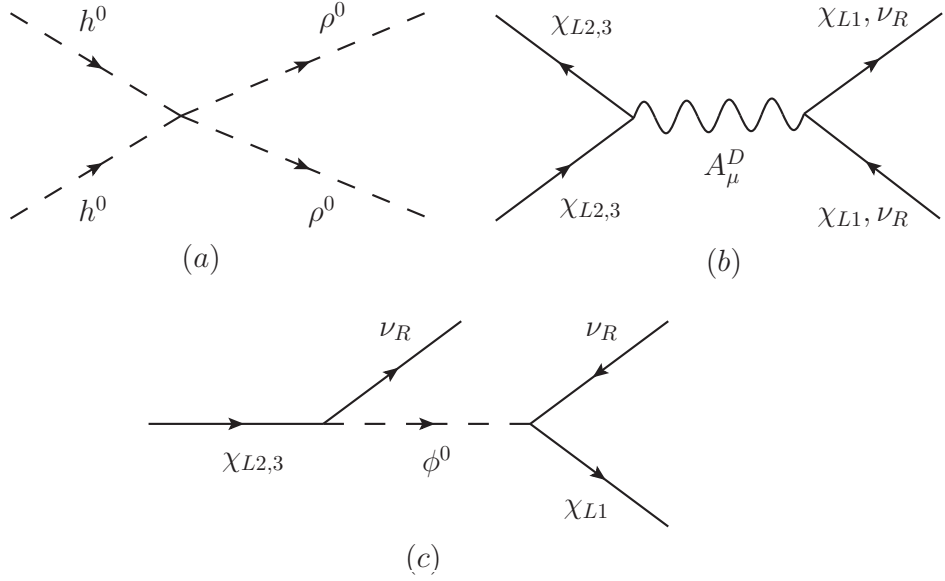}
 \caption{(a) A pair of the SM Higgs bosons annihilating into a pair of the dark scalar bosons, by which the $\nu SM$ sector can communicate with the dark sector, but these two sectors are isolated from each other at $T<T_{D}$. (b) The symmetric part of $\chi_{2,3}$ was exhausted by the annihilation process $\chi_{2,3}+\widetilde{\chi}_{2,3}\rightarrow\chi_{1}+\widetilde{\chi}_{1}/\nu_{R}+\widetilde{\nu_{R}}$\,, but its asymmetric part can survive without any loss, eventually the surviving $\chi_{2,3}$ become the CDM. (c) The CDM $\chi_{2,3}$ can very slowly decay by $\chi_{2,3}\rightarrow\nu_{R}+\widetilde{\nu_{R}}+\chi_{1}$, but its lifetime is far larger than the present universe age.}
\end{figure}
  The evolution processes inside the dark sector are similar to ones inside the $\nu SM$ sector. In the light of the last line in Eq. (8) and the third line in Eq. (9), the massive $\rho^{0}$ and $A_{\mu}^{D}$ can decay into a pair of lighter $\chi_{i}$ or $\nu_{R}$, namely
\begin{alignat}{1}
 &\rho^{0}\rightarrow A^{D}+A^{D},\hspace{0.5cm}
  \rho^{0}\rightarrow\chi_{i}+\widetilde{\chi}_{i}\,,\nonumber\\
 &A_{\mu}^{D}\rightarrow \nu_{R}+\widetilde{\nu_{R}}\,,\hspace{0.5cm}
  A_{\mu}^{D}\rightarrow \chi_{Li}+\widetilde{\chi_{L}}_{i}\,.
 \end{alignat}
  However, the massive $\chi_{2,3}$ can almost not decay, therefore it is very stable particle. In fact, the only decay channel of $\chi_{2,3}$ is $\chi_{2,3}\rightarrow\nu_{R}+\widetilde{\nu_{R}}+\chi_{1}$ via the $\phi^{0}$ mediation, as shown the (c) diagram in Fig. 3, but its decay width is so small due to the $M^{-4}_{\phi^{0}}$ suppression that the $\chi_{2,3}$ lifetime is far larger than the present universe age about $1.38\times10^{10}$ year \cite{1}. The detailed calculation is as follows,
\begin{alignat}{1}
 &\tau_{\chi_{2,3}}^{-1}=\Gamma[\chi_{2,3}\rightarrow\nu_{R}+\widetilde{\nu_{R}}+\chi_{1}]=
  \frac{m^{5}_{\chi_{i}}(Y_{3}^{\dagger}Y_{3})_{ii}(Y_{3}^{\dagger}Y_{3})_{11}}
  {768(2\pi)^{3}M^{4}_{\phi^{0}}}\,,\nonumber\\
 \Longrightarrow &\tau_{\chi_{2,3}}\gtrsim1.4\times10^{17}\:\mathrm{year}\;
  \mbox{for}\:M_{\phi^{0}}\sim 10^{11}\:\mathrm{GeV},
\end{alignat}
  where $i=2,3$ and $m_{\chi_{2,3}}\approx3.4$ GeV which will be given by the following Eq. (20). Therefore, the dark fermion $\chi_{2,3}$ is indeed as extremely stable as the SM proton and electron, it exactly acts as the CDM.

  $\chi_{2,3}+\widetilde{\chi}_{2,3}$ can annihilate into $\chi_{1}+\widetilde{\chi}_{1}$ or $\nu_{R}+\widetilde{\nu_{R}}$ via the dark photon s-channel mediation, as shown the (b) diagram in Fig. 3, by contrast, $\chi_{2,3}+\chi_{2,3}$ or $\widetilde{\chi}_{2,3}+\widetilde{\chi}_{2,3}$ is only an elastic scattering via the dark photon t-channel mediation. However, the process cross-sections in these two cases are approximately equal. When the $\chi_{2,3}$ annihilate rate is smaller than the universe expansion rate, the annihilating process is frozen, thus $\chi_{2,3}$ is non-relativistic decoupling, at the same time, $\chi_{1}$ and $\nu_{R}$ is relativistic decoupling. The freeze-out temperature is calculated by the following relations,
\begin{alignat}{1}
 &\Gamma[\chi_{L2,3}+\widetilde{\chi_{L}}_{2,3}\rightarrow
  \chi_{L1}+\widetilde{\chi_{L}}_{1}\:or\:\nu_{R}+\widetilde{\nu_{R}}]
  =\langle\sigma v_{r}\rangle n_{\chi_{2,3}}(T_{f})\lesssim H(T_{f})\,,\nonumber\\
 &\langle\sigma v_{r}\rangle=\frac{\alpha_{D}}{2v_{\phi}^{2}}\frac{r}{(1-4r)^{2}}
  (1+\frac{1-2r}{1-4r}\,\frac{6}{x})\,,\hspace{0.3cm} n_{\chi_{2,3}}=T_{f}^{3}\left(\frac{x}{2\pi}\right)^{\frac{3}{2}}e^{-x}\,,\nonumber\\
 &\alpha_{D}=\frac{e_{D}^{2}}{4\pi}\sim10^{-2}\,,\hspace{0.3cm}
  r=\frac{m^{2}_{\chi_{2,3}}}{M^{2}_{A^{D}}}\sim1,\hspace{0.3cm}
  x=\frac{m_{\chi_{2,3}}}{T_{f}}\,,\nonumber\\
 \Longrightarrow &x\approx 40.7+ln[m_{\chi_{2,3}}\langle\sigma v_{r}\rangle
  \sqrt{\frac{x}{g_{*}(T_{f})}}\,]\,,\nonumber\\
 \Longrightarrow &\langle\sigma v_{r}\rangle\sim 10^{-6}\:\mathrm{GeV^{-2}}\;
  \mbox{for}\:v_{\phi}\sim 10\:\mathrm{GeV},\nonumber\\
 &T_{f}\approx0.12\:\mathrm{GeV}\;\mbox{for}\:m_{\chi_{2,3}}\approx3.4\:\mathrm{GeV},
\end{alignat}
  where $g_{*}(T_{f})=19.5$ since the relativistic states only includes $\chi_{1},\nu_{R},\nu_{L},e,\gamma$ at $T_{f}\approx0.12$ GeV, and $m_{\chi_{2,3}}\approx3.4$ GeV will be given by the following Eq. (20). The thermal average $\langle\sigma v_{r}\rangle\sim 10^{-6}\:\mathrm{GeV^{-2}}$ is four orders of magnitude larger than the usual weak interaction cross-section of $\sim 10^{-10}\:\mathrm{GeV^{-2}}$ because $M^{2}_{A^{D}}$ is much smaller than $M^{2}_{Z}$ in the gauge propagator. As a result, the symmetric part of $\chi_{2,3}$ and $\widetilde{\chi}_{2,3}$ is sufficiently annihilated and exhausted, but the asymmetric part of $\chi_{2,3}$ and $\widetilde{\chi}_{2,3}$ which arise from the inflaton $\phi^{0}$ decay, namely $\eta_{\chi}$ in Eq. (15), can survive without any loss. This picture is analogous to one in the $\nu SM$ sector in which the symmetric part of the baryon is annihilated via the photon and only its asymmetric part can survive. On the other hand, the elastic scattering among the surviving $\chi_{2,3}$ is also frozen at the temperature of $\sim T_{f}$. Therefore, at $T<T_{f}$ the surviving $\chi_{2,3}$ are actually free particles except for the gravitational influence, eventually, they become the CDM in the present universe.

  From the above discussions, we can see that the evolutions in the dark sector are very similar to ones in the $\nu SM$ sector.  At the present day, the asymmetric baryon exists in the $\nu SM$ sector while the asymmetric CDM $\chi_{2,3}$ inhabits in the dark sector. The density abundances of the baryon and the CDM $\chi_{2,3}$ are given by the following equations,
\begin{alignat}{1}
 &\Omega_{B}h^{2}=\frac{n_{\gamma}(T_{0})\eta_{B}m_{p}}{\rho_{c}}\,h^{2}\,,\hspace{0.3cm}
  \Omega_{CDM}h^{2}=\frac{n_{\gamma}(T_{0})\frac{\eta_{\chi}}{3}\sum\limits_{i}m_{\chi_{i}}}{\rho_{c}}\,h^{2}\,,\nonumber\\
 \Longrightarrow &\frac{\Omega_{B}}{\Omega_{CDM}}=
  \frac{3\eta_{B}m_{p}}{\eta_{\chi}\sum\limits_{i}m_{\chi_{i}}}\approx0.188\,,\nonumber\\
 \Longrightarrow &m_{\chi_{2}}=m_{\chi_{3}}\approx 3.4\:\mathrm{GeV},
\end{alignat}
  where $\rho_{c}=1.054\times10^{-5}h^{2}\:GeV/cm^{3}$ is the critical density, $n_{\gamma}(T_{0})\approx411/cm^{3}$ is the CMB photon number density at the present temperature $T_{0}\approx2.73$ K \cite{1}, $m_{p}=0.938$ GeV is the proton mass, and $0.188$ is the current ratio of the baryon density to the CDM one from multiple experiments \cite{1,23}. By use of $\frac{\eta_{\chi}}{\eta_{B}}=\frac{79}{36}$ in Eq. (15), we can obtain $m_{\chi_{2,3}}\approx3.4$ GeV, thus the CDM $\chi_{2,3}$ mass is accurately determined by the model. In conclusion, we have demonstrated that the physical natures of $\chi_{2,3}$ and its relic abundance are very well consistent with the requirements for the CDM, so $\chi_{2,3}$ is indeed a desirable candidate of the CDM.

  In the model, there are in all four relativistic decoupling particles, i.e., $\gamma,\nu_{L}$ in the $\nu SM$ sector and $\nu_{R},\chi_{1}$ in the dark sector. At $T<T_{D}\sim(7-15)$ GeV, the $\nu SM$ sector and the dark one are isolated from each other, therefore the entropy in each sector is separately conserved. In the $\nu SM$ sector, it is well known that the massless $\gamma$ has become the CMB with $T_{0}\approx2.73$ K and the decoupling $\nu_{L}$ has the effective temperature of $T_{\nu_{L}}\approx1.95$ K in the present universe. In the dark sector, $\nu_{R}$ and $\chi_{1}$ were together relativistic decoupling at the same freeze-out temperature $T_{f}\approx0.12$ GeV, see Eq. (19), therefore they has the same effective temperature, namely $T_{\chi_{1}}=T_{\nu_{R}}$. Eventually, $\nu_{L}$ and $\nu_{R}$ with the Sub-eV mass become the hot dark matter, while the massless $\chi_{1}$ becomes a dark background radiation. We can derive the same effective temperature of $\nu_{R}$ and $\chi_{1}$ by the entropy conservation in the dark sector, the detailed calculations are as follows,
\begin{alignat}{1}
 &\frac{s^{Dark}(T_{D})a^{3}(T_{D})}{s^{SM}(T_{D})a^{3}(T_{D})}
  =\frac{s^{Dark}(T_{0})a^{3}(T_{0})}{s^{SM}(T_{0})a^{3}(T_{0})}\,,\nonumber\\
 \Longrightarrow&\frac{g_{*}^{Dark}(T_{D})}{g_{*}^{SM}(T_{D})}=\frac{g_{*}^{Dark}(T_{0})}{g_{*}^{SM}(T_{0})}=
  \frac{g_{*}^{\chi_{1}}(\frac{T_{\chi_{1}}}{T_{0}})^{3}+g_{*}^{\nu_{R}}(\frac{T_{\nu_{R}}}{T_{0}})^{3}}
  {g_{*}^{\gamma}+g_{*}^{\nu_{L}}(\frac{T_{\nu_{L}}}{T_{0}})^{3}}\,,\nonumber\\
 &g_{*}^{Dark}(T_{D})=g_{*}^{\chi_{1}}+g_{*}^{\nu_{R}}=8.75\,,\nonumber\\
 &g_{*}^{SM}(T_{D})=g_{*}^{\gamma}+g_{*}^{gluon}+g_{*}^{u,d,s}+g_{*}^{e,\mu}+g_{*}^{\nu_{L}}=61.75\,,\nonumber\\
 &(\frac{T_{\nu_{L}}}{T_{0}})^{3}=\frac{4}{11}\,,\hspace{0.5cm}T_{\chi_{1}}=T_{\nu_{R}}\,,\nonumber\\
 \Longrightarrow&(\frac{T_{\chi_{1}}}{T_{0}})^{3}=(\frac{T_{\nu_{R}}}{T_{0}})^{3}=0.0633\,,\nonumber\\
 \Longrightarrow&T_{\chi_{1}}=T_{\nu_{R}}\approx 1.1\:\mathrm{K}<T_{\nu_{L}}\approx 1.95\:\mathrm{K}
  <T_{0}\approx2.73\:\mathrm{K}\,,
\end{alignat}
  where $a(T)$ is the scale factor of the universe expansion. Finally, the present density abundances of all kinds of the relativistic decoupling particles are given by the following relations,
\begin{alignat}{1}
 &\rho_{\gamma}(T_{0})=\frac{\pi^{2}}{30}g_{*}^{\gamma}T_{0}^{4}\,,\hspace{0.3cm}
  \Omega_{\gamma}h^{2}=\frac{\rho_{\gamma}(T_{0})}{\rho_{c}}\,h^{2}\approx2.5\times10^{-5}\,,\nonumber\\
 &\rho_{\chi_{1}}(T_{0})=\frac{\pi^{2}}{30}g_{*}^{\chi_{1}}(\frac{T_{\chi_{1}}}{T_{0}})^{4}T_{0}^{4}\,,\hspace{0.3cm}
  \Omega_{\chi_{1}}h^{2}=\frac{\rho_{\chi_{1}}(T_{0})}{\rho_{c}}\,h^{2}\approx1.1\times10^{-6}\,,\nonumber\\
 &\rho_{\nu_{L1}}(T_{0})=\frac{\pi^{2}}{30}g_{*}^{\nu_{L1}}(\frac{T_{\nu_{L}}}{T_{0}})^{4}T_{0}^{4}\,,\hspace{0.3cm}
  \Omega_{\nu_{L1}}h^{2}=\frac{\rho_{\nu_{L1}}(T_{0})}{\rho_{c}}\,h^{2}\approx5.6\times10^{-6}\,,\nonumber\\
 &\rho_{\nu_{R1}}(T_{0})=\frac{\pi^{2}}{30}g_{*}^{\nu_{R1}}(\frac{T_{\nu_{R}}}{T_{0}})^{4}T_{0}^{4}\,,\hspace{0.3cm}
  \Omega_{\nu_{R1}}h^{2}=\frac{\rho_{\nu_{R1}}(T_{0})}{\rho_{c}}\,h^{2}\approx5.5\times10^{-7}\,,\nonumber\\
 &n_{\nu_{L}}(T_{0})=\frac{3}{4}(\frac{T_{\nu_{L}}}{T_{0}})^{3}n_{\gamma}(T_{0})\approx112/cm^{3}\,,\hspace{0.3cm}
  \Omega_{\nu_{L2,3}}h^{2}=\frac{n_{\nu_{L}}(T_{0})\sum\limits_{i}m_{\nu_{i}}}{\rho_{c}}\,h^{2}\approx6.3\times10^{-4}\,,\nonumber\\
 &n_{\nu_{R}}(T_{0})=\frac{3}{4}(\frac{T_{\nu_{R}}}{T_{0}})^{3}n_{\gamma}(T_{0})\approx19.5/cm^{3}\,,\hspace{0.3cm}
  \Omega_{\nu_{R2,3}}h^{2}=\frac{n_{\nu_{R}}(T_{0})\sum\limits_{i}m_{\nu_{i}}}{\rho_{c}}\,h^{2}\approx1.1\times10^{-4}\,,
\end{alignat}
  where $\gamma,\chi_{1},\nu_{1}$ are massless and $\nu_{2,3}$ are massive. These results of Eq. (22) are very well consistent with the current density budgets of all kinds of matters in the universe, one can refer to the review of cosmological parameters in \cite{1}.

\vspace{0.6cm}
\noindent\textbf{V. Numerical Results and Model Test}

\vspace{0.3cm}
  We now summarize the model by some concrete numerical results. All kinds of the parameters of the SM have essentially been fixed by the current experimental data \cite{1}, while several key parameters in the dark sector can be determined very well by the current data of the tiny neutrino mass, the baryon asymmetry, and the CDM abundance. For simplicity, we can demonstrate the model by the following set of typical values in the parameter space,
\begin{alignat}{1}
 &v_{\Phi}=10^{16}\:\mathrm{GeV},\hspace{0.3cm} v_{H}=246\:\mathrm{GeV},\hspace{0.3cm}
  v_{\phi}=10\:\mathrm{GeV},\nonumber\\
 &M_{\phi^{0}}=2\times10^{11}\:\mathrm{GeV},\hspace{0.3cm}
  \frac{M_{\phi^{0}}}{m_{N_{1}}}=10^{-2}\,,\hspace{0.3cm}
  m_{\chi_{2,3}}=3.4\:\mathrm{GeV},\nonumber\\
 &\frac{Im[Tr[(Y_{1}\frac{m_{N_{1}}}{M_{N}}Y_{2})Y_{3}Y_{\chi}^{\dagger}
  (Y_{1}\frac{m_{N_{1}}}{M_{N}}Y_{3}^{*})^{\dagger}]]}
  {Tr[Y_{3}Y_{3}^{\dagger}]}=-1,\nonumber\\
 &Eigenvalues[(Y_{1}\frac{m_{N_{1}}}{M_{N}}Y_{2})(Y_{1}\frac{m_{N_{1}}}{M_{N}}Y_{2})^{\dagger}]
  =(0,0.0195,0.673)\,.
\end{alignat}
  $v_{\Phi}$ is just at the scale of $\Lambda_{GUT}$, $v_{H}$ is fixed by the SM physics, and $v_{\phi}$ is determined by the dark sector physics. $M_{\phi^{0}}$ is naturally close to the reheating temperature of $\sim10^{12}$ GeV in the popular theories of the universe inflation and reheating. $\frac{M_{\phi^{0}}}{m_{N_{1}}}$ is determined by fitting the baryon asymmetry $\eta_{B}$, while $m_{\chi_{2,3}}$ is fixed by fitting the CDM abundance $\Omega_{CDM}h^{2}$. Because $Y_{1},Y_{2},Y_{3},Y_{\chi},\frac{m_{N_{1}}}{M_{N}}$ are all $\sim\mathcal{O}(1)$, these matrix operation results in the last two lines in Eq. (23) are all consistent and reasonable, they are taken as input parameter values. In short, all of the parameter values in Eq. (23) is completely in accordance with the model requirements and the previous discussions.

  Substitute Eq. (23) into the equations of (9),(12),(14),(15),(20), then we can correctly reproduce the desired results,
\begin{alignat}{1}
 &\triangle m_{21}\approx 7.53\times10^{-5}\:\mathrm{eV^{2}}\,,\hspace{0.3cm}
  \triangle m_{32}\approx 2.52\times10^{-3}\:\mathrm{eV^{2}}\,,\nonumber\\
 &\eta_{B}\approx6.1\times10^{-10}\,,\hspace{0.3cm}\Omega_{B}h^{2}\approx0.0222\,,\nonumber\\
 &\eta_{\chi}\approx13.3\times10^{-10}\,,\hspace{0.3cm}\Omega_{CDM}h^{2}\approx0.118\,,
\end{alignat}
  where $\triangle m_{ij}=m^{2}_{\nu_{i}}-m^{2}_{\nu_{j}}$ and $m_{\nu_{1}}=0$. Obviously, Eq. (24) are very well in agreement with the current experimental data \cite{1}. Furthermore, we can fit the full experimental data of the neutrino masses and mixing angles if we choose some suitable texture of the matrix $Y_{1}\frac{m_{N_{1}}}{M_{N}}Y_{2}$, but we give up this discussion to limit the length of this paper. In conclusion, only by use of these simple and natural parameters without any fine-tuning in Eq. (23), the model can completely account for the three puzzles of the neutrino mass, the baryon asymmetry, and the CDM, therefore, this fully demonstrates that the model is very successful and believable.

  In the end, we simply discuss several methods for this model test. At the low energy, the dark sector has essentially decoupled from the $\nu SM$ sector, the connection between these two sectors is only feeble, therefore any detections for the dark sector will be very difficult. On the basis of those couplings of the model, Fig. 4 shows three feasible approaches by which we can probe the dark sector and test the model. The (a) diagram in Fig. 4 shows that an $\alpha$-flavour ${\nu_{L\alpha}}$ in the $\nu SM$ sector is scattering with a CDM $\chi_{2,3}$ via the $\rho^{0}$ mediation, and then this ${\nu_{L\alpha}}$ is converted into a $\beta$-flavour $\nu_{R\beta}$ in the dark sector. The scattering cross-section is given by
\begin{alignat}{1}
 &\sum\limits_{\beta}\sigma[\nu_{L\alpha}+\chi_{2,3}\rightarrow \nu_{R\beta}+\chi_{2,3}]=
  \frac{(M_{\nu}M_{\nu}^{\dagger})_{\alpha\alpha}}{4\pi v^{4}_{\phi}}
  (\frac{m_{\chi_{2,3}}}{M_{\rho^{0}}})^{4}
  (\frac{E_{\nu_{L\alpha}}}{m_{\chi_{2,3}}+2E_{\nu_{L\alpha}}})^{2}\,,\nonumber\\
 &M_{\nu}M_{\nu}^{\dagger}=U_{\nu}\,Diag(m^{2}_{\nu_{1}},m^{2}_{\nu_{2}},m^{2}_{\nu_{3}})U_{\nu}^{\dagger}\,,
\end{alignat}
  where $E_{\nu_{L\alpha}}$ is the incident energy of the $\alpha$-flavour $\nu_{L\alpha}$ and we assume $E_{\nu_{L\alpha}}<m_{\chi_{2,3}}$, and $U_{\nu}$ is namely the mixing matrix of the SM $\nu_{L}$. For the electronic neutrino with $E_{\nu_{Le}}=1$ GeV, we can roughly estimate $\sigma\approx2\times10^{-30}/GeV^{2}$ provided $M_{\rho^{0}}=6$ GeV. Although this cross-section is very small, it is possible to find $\nu_{L}\rightarrow\nu_{R}$ by the cosmic neutrino detection, for instance, we can detect the $\nu_{L}$ stream emitted by a distant supernova, a tiny part of the $\nu_{L}$ stream can however scatter with the CDM $\chi_{2,3}$ in the galactic halo, thus they are converted into the dark $\nu_{R}$ before they can arrive to the earth, therefore the $\nu_{L}$ number which can eventually arrive to the earth is certainly less than the expected value. This detection is very similar to the detection for the flavour conversion of the solar neutrino. This method can not only confirm the CDM $\chi_{2,3}$, but also directly shed light on the neutrino mass origin, namely the Dirac-type seesaw mechanism employed by the model.

\begin{figure}
 \centering
 \includegraphics[totalheight=8cm]{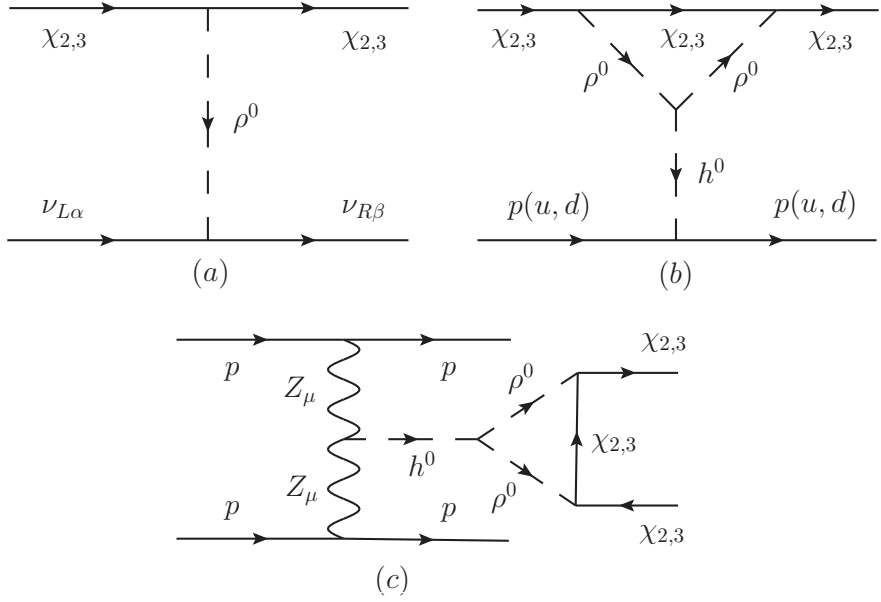}
 \caption{(a) The SM $\nu_{L}$ can be converted into the dark $\nu_{R}$ through its scattering with the CDM $\chi_{2,3}$\,. (b) The CDM $\chi_{2,3}$ can be detected by its elastic scattering with the nucleon. (c) The CDM $\chi_{2,3}$ and the other dark particles can be searched at the LHC.}
\end{figure}
  The (b) diagram in Fig. 4 is an elastic scattering of the CDM $\chi_{2,3}$ off the nuclei. This process is just a goal for many underground detectors which are endeavouring to search some evidences of the CDM \cite{5,24}. Its scattering cross-section is estimated as $\lesssim10^{-20}/GeV^{2}$, obviously, it is far below the current experiment limits of the direct detection for the CDM. Although we can not yet detect it under the present conditions, it is promising to reach this goal in the near future.

  At the present LHC \cite{25}, we have an opportunity to search the CDM $\chi_{2,3}$ and the other dark particles.  The (c) diagram in Fig. 4 shows a relevant process. Of course, this search is very difficult because its cross-section is too small, on the other hand, the dark particles can escape from the detectors. However, some planned colliders such as CEPC and ILC have some better potentials to reach this goal \cite{26}.

  The scientific significance of the above-mentioned experimental searches are beyond all doubt. Although it will be very large challenges to actualize them, it is not impossible. In the near future, it is very possible that we shall be able to probe the dark sector physics beyond the SM and open a window of the dark universe.

\vspace{0.6cm}
\noindent\textbf{VI. Conclusions}

\vspace{0.3cm}
  In summary, I suggest a new extension of the SM of particle physics. This model introduces the dark sector with the dark $SU(2)_{D}\times U(1)_{D'}$ gauge symmetry. The new particles inhabiting in the dark sector are all singlets under the SM groups. At the GUT scale, the global $B-L$ conservation and the dark gauge symmetry are together broken by the dark doublet scalar vacuum expectation $\langle\Phi\rangle$, but the global $B-L-L^{D}$ number instead of the $B-L$ number and the dark $Q^{D}$ charge are conversed as two residual symmetries. At the low energy, $\langle H\rangle$ breaks the electroweak symmetry in the SM sector, while $\langle\phi^{-}\rangle$ violates the $Q^{D}$ charge conservation in the dark sector, but the global $B-L-L^{D}$ conservation is still kept in these two sectors. The SM particle masses and the dark particle ones respectively arise from these symmetry breakings. In particular, the SM $\nu_{L}$ and the dark $\nu_{R}$ are combined to generate a tiny Dirac neutrino mass by the Dirac-type seesaw mechanism.

  In the model, the neutral real scalar field $\phi^{0}$ in the dark sector plays a role of the inflaton. Its two-body decay can provide the universe inflation and reheating, simultaneously, its three-body decay can generate the $B-L$ asymmetry in the SM sector and the $L^{D}$ asymmetry in the dark sector, and then they respectively lead to the baryon asymmetry and the asymmetric CDM $\chi_{2,3}$. The dark Dirac fermion $\chi_{2,3}$ has $m_{\chi_{2,3}}\approx3.4$ GeV, and it is extremely stable due to the $Z_{4}$ symmetry protection, in fact, its lifetime is far larger than the universe age. The symmetric part of $\chi_{2,3}$ can completely annihilate into a pair of the massless $\chi_{1}$ or the light $\nu_{R}$ via the dark photon mediation, but its asymmetric part can survive without loss, eventually, the surviving $\chi_{2,3}$ become the CDM in the present universe. The evolution picture in the dark sector is similar to one in the SM sector. Because the $\chi_{2,3}$ asymmetry and the baryon one together originate from the inflaton $\phi^{0}$ decay, the relic abundance of the CDM $\chi_{2,3}$ is similar in size to the baryon abundance in the current universe.

  In short, the model can completely account for the common origin of the tiny neutrino mass, the baryon asymmetry and the asymmetric CDM only by a few simple and natural parameters. In addition, the model gives some interesting predictions for the dark sector physics, for instance, the CDM $\chi_{2,3}$ mass is accurately $3.4$ GeV, there are the dark photon and the dark neutral scalar boson whose masses are several GeVs, and the dark background radiation formed by the massless $\chi_{1}$. Finally, I give the three feasible approaches to test the model by means of the supernova neutrino physics, the underground detectors, and the TeV collider experiments. In the near future, it is very possible that we shall be able to probe the dark sector physics beyond the SM and open a window of the dark universe.

\vspace{0.6cm}
 \noindent\textbf{Acknowledgements}

\vspace{0.3cm}
  I would like to thank my wife for her great helps. This research is supported by the Fundamental Research Funds for the Central Universities Grant No. WY2030040065.

\vspace{0.6cm}
\noindent\textbf{Appendix A}

\vspace{0.3cm}
  After the inflaton $\phi^{0}$ decay is completed, the dark plasma consists of $A_{\mu}^{D},\phi^{\mp},\chi_{L}^{-},\chi_{R}^{0},\nu_{R}^{-}$. Because the total dark charge is conserved, and the $Y_{\chi}$ and $Y_{3}$ Yukawa interactions in the third and fourth rows in Eq. (9) are still in equilibrium, the chemical potentials of these dark particle states need satisfy the relations as follows,
\begin{alignat}{1}
 &\sum\limits_{i}(-\mu_{\chi_{Li}}-\mu_{\nu_{Ri}}-2\mu_{\phi^{-}}\frac{2}{N_{f}})=0,\hspace{0.3cm}
  \mu_{\chi_{Ri}}-\mu_{\chi_{Lj}}+\mu_{\phi^{-}}=0,\hspace{0.3cm}
  \mu_{\nu_{Ri}}-\mu_{\chi_{Lj}}=0,\nonumber\\
 &\Longrightarrow \mu_{\chi_{Li}}=\mu_{\nu_{Ri}}\,,\hspace{0.3cm}
  \mu_{\chi_{Ri}}=\frac{2+N_{f}}{2}\mu_{\nu_{Ri}}\,,\hspace{0.3cm} \mu_{\phi^{-}}=-\frac{1}{2}\sum\limits_{i}\mu_{\nu_{Ri}}\,.
\end{alignat}
  From the above equations, the asymmetries of the dark leptons are thus solved out,
\begin{alignat}{1}
 &Y_{L^{D}}=k\sum\limits_{i}(\mu_{\chi_{Li}}+\mu_{\chi_{Ri}}+\mu_{\nu_{Ri}})=
  k\frac{6+N_{f}}{2}\sum\limits_{i}\mu_{\nu_{Ri}}\,,\nonumber\\
 &Y_{\chi}=k\sum\limits_{i}(\mu_{\chi_{Li}}+\mu_{\chi_{Ri}})=
  k\frac{4+N_{f}}{2}\sum\limits_{i}\mu_{\nu_{Ri}}\,,\hspace{0.3cm}
  Y_{\nu_{R}}=k\sum\limits_{i}\mu_{\nu_{Ri}}\,,\nonumber\\
 \Longrightarrow &Y_{\chi}=c_{\chi}Y_{L^{D}},\hspace{0.3cm}
  Y_{\nu_{R}}=(1-c_{\chi})Y_{L^{D}},\hspace{0.3cm} c_{\chi}=\frac{4+N_{f}}{6+N_{f}}\xrightarrow{N_{f}=3}\frac{7}{9}\,,
\end{alignat}
  where $k=\frac{T^{2}}{6s}$\,.

\vspace{0.6cm}
\noindent\textbf{Appendix B}

\vspace{0.3cm}
  At the temperature of $T_{0}\approx2.73$ K, the ratio of the total entropy density to the photon number density is calculated by
\begin{alignat}{1}
 &n_{\gamma}(T_{0})=\frac{1.2}{\pi^{2}}2T_{0}^{3}\,,\hspace{0.3cm}s(T_{0})=\frac{2\pi^{2}}{45}T_{0}^{3}
  \left(g_{*}^{\gamma}+g_{*}^{\nu_{L}}(\frac{T_{\nu_{L}}}{T_{0}})^{3}+g_{*}^{\nu_{R}}(\frac{T_{\nu_{R}}}{T_{0}})^{3}
  +g_{*}^{\chi_{1}}(\frac{T_{\chi_{1}}}{T_{0}})^{3}\right),\nonumber\\
 &(\frac{T_{\nu_{L}}}{T_{0}})^{3}=\frac{4}{11}\,,\hspace{0.3cm}
  (\frac{T_{\nu_{R}}}{T_{0}})^{3}=(\frac{T_{\chi_{1}}}{T_{0}})^{3}=0.0633\,,\nonumber\\
 &\Longrightarrow \frac{s(T_{0})}{n_{\gamma}(T_{0})}=8.05\,,
\end{alignat}
  where $(\frac{T_{\nu_{R}}}{T_{0}})^{3}=(\frac{T_{\chi_{1}}}{T_{0}})^{3}=0.0633$ has been given in Eq. (21).


\vspace{0.3cm}
\begin{thebibliography}{99}
\bibitem{1}
  M. Tanabashi, \emph{et al.} (Particle Data Group), Phys. Rev. D 98, 030001 (2018).
\bibitem{2}
  S. F King, A. Merle, S. Morisi, Y. Shimizu and M. Tanimoto, New J. Phys. 16, 045018 (2014);
  G. Altarelli, Int. J. Mod. Phys. A 29, 1444002 (2014);
  R. N. Mohapatra, \emph{et al.}, Rep. Prog. Phys. 70, 1757 (2007).
\bibitem{3}
  L. Canetti, M. Drewes and M. Shaposhnikov, New J. Phys. 14, 095012 (2012);
  M. Dine and A. Kusenko, Rev. Mod. Phys. 76, 1 (2004).
\bibitem{4}
  G. B. Gelmini, arXiv:1502.01320;
  V. Lukovic, P. Cabella and N. Vittorio, Int. J. Mod. Phys. A 29, 1443001 (2014);
  G. Bertone, Particle Dark Matter (Cambridge University Press, 2010).
\bibitem{5}
  F. Mayeta, \emph{et al}, Phys. Reps. 627, 1 (2016).
\bibitem{6}
  M. Gell-Mann, P. Ramond, R. Slansky, in Supergravity, eds. P. van Niewenhuizen and D. Z. Freeman (North-Holland, Amsterdam, 1979);
  T. Yanagida, in Proc. of the Workshop on Unified Theory and Baryon Number in the Universe, eds. O. Sawada and A. Sugamoto (Tsukuba, Japan, 1979);
  R. N. Mohapatra, G. Senjanovic, Phys. Rev. Lett. 44, 912 (1980).
\bibitem{7}
  A. de Gouvea, Annu. Rev. Nucl. Part. Sci. 66, 197 (2016).
\bibitem{8}
  W. Buchmuller, R. D. Peccei and T. Yanagida, Annu. Rev. Nucl. Part. Sci. 55, 311 (2005);
  S. Davidson, E. Nardi and Y. Nir, Phys. Reps. 466, 105 (2008).
\bibitem{9}
  D. E. Morrissey and M. J. Ramsey-Musolf, New J. Phys. 14, 125003 (2012);
  J. M. Cline, arXiv:hep-ph/0609145.
\bibitem{10}
  A. Kusenko, Phys. Reps. 481, 1 (2009);
  K. N. Abazajian, Phys. Reps. 711-712, 1 (2017);
  K. N. Abazajiana, \emph{et al.}, arXiv:1204.5379.
\bibitem{11}
  G. Jungman, M. Kamionkowski, and K. Griest, Phys. Reps. 267, 195 (1996).
\bibitem{12}
  David J.E. Marsh, Phys. Reps. 643, 1 (2016).
\bibitem{13}
  M. Fukugita and T. Yanagida, Phys. Lett. B 174, 45 (1986);
  W. Buchmuller, Nucl. Phys. B (Proc. Suppl.) 235–236, 329 (2013).
\bibitem{14}
  Ernest Ma, Utpal Sarkar, Phys.Rev.Lett. 80, 5716 (1998).
\bibitem{15}
  K. M. Zurek, Phys. Reps. 537, 91 (2014);
  B. Fornal, Y. Shirman, Tim M. P. Tait, and J. R. West, Phys. Rev. D 96, 035001 (2017);
  J. Sakstein, M. Trodden, Phys. Lett. B774, 183 (2017).
\bibitem{16a}
  J. D. Clarke and R. R. Volkas, Phys. Rev. D 93, 035001 (2016);
  S. Kashiwase, D. Suematsu, Phys. Rev. D 86, 053001 (2012);
  M. Aoki, S. Kanemura, O. Seto, Phys. Rev. D 80, 033007 (2009).
\bibitem{16b}
  W. M. Yang, J. High Energy Phys 03, 144 (2018);
  W. M. Yang, Phys. Lett. B 762, 138 (2016);
  W. M. Yang, Nucl. Phys. B 885, 505 (2014);
  W. M. Yang, Phys. Rev. D 87, 095003 (2013).
\bibitem{16c}
  T. Asaka and M. Shaposhnikov, Phys. Lett. B 620, 17 (2005).
\bibitem{17}
  A. Linde, Inflationary Cosmology, Lect. Notes Phys. 738, 1–54 (2008).
\bibitem{18}
  R. Allahverdi, R. Brandenberger, F.-Yan C.-Racine, and A. Mazumdar, Annu. Rev. Nucl. Part. Sci. 60, 27 (2010).
\bibitem{19}
  A. D. Sakharov, Pisma Zh. Eksp. Teor. Fiz. 5, 32 (1967), JETP Lett. 5, 24 (1967),
  Sov. Phys. Usp. 34, 392 (1991), Usp. Fiz. Nauk 161, 61 (1991).
\bibitem{20}
  E. W. Kolb and M. S. Turner, The Early universe, Front. Phys. 69, 1 (1990);
  D. S. Gorbunov and V. A. Rubakov, Introduction to The Theory of The Early Universe: Hot Big Bang Theory (World Scientific Publishing Co. Pte. Ltd, 2011).
\bibitem{21}
  V. A. Kuzmin, V. A. Rubakov, M. A. Shaposhnikov, Phys. Lett. B 155, 36 (1985).
\bibitem{22}
  E. Komatsu, \emph{et al} (WMAP Collaboration), Astrophys. J. Suppl. 192, 18 (2011).
\bibitem{23}
  Planck Collab. 2015 Results XIII, Astron. \& Astrophys., arXiv:1502.01589v2.
\bibitem{24}
  M. Klasen, M. Pohl, G. Sigl, Progress in Particle and Nuclear Physics 85, 1-32 (2015);
  R. W. Schnee, arXiv:1101.5205.
\bibitem{25}
  D. E. Morrissey, T. Plehn, T. M. P. Tait, Phys. Reps. 515, 1 (2012).
\bibitem{26}
  CEPC-SPPC Study Group, CEPC-SPPC Preliminary Conceptual Design Report. 1. Physics and Detector, (2015);
  B. Barish and J. E. Brau, Int. J. Mod. Phys. A 28, 1330039 (2013).

\end{thebibliography}
\end{document}